*Article*

# Examination of Cybersickness in Virtual Reality: The Role of Individual Differences, Effects on Cognitive Functions & Motor Skills, and Intensity Differences During and After Immersion.

**Panagiotis Kourtesis[1-3]\*, Agapi Papadopoulou[1], and Petros Roussos[1]**

[1]   Department of Psychology, National and Kapodistrian University of Athens, Athens, Greece
[2]   Department of Psychology, American College of Greece, Athens, Greece
[3]   Department of Psychology, University of Edinburgh, Edinburgh, United Kingdom
\*   Correspondence: pkourtesis@psych.uoa.gr

**Abstract:** Background: Given that VR is applied in multiple domains, understanding the effects of cybersickness on human cognition and motor skills and the factors contributing to cybersickness gains urgency. This study aimed to explore the predictors of cybersickness and its interplay with cognitive and motor skills. Methods: 30 participants, 20-45 years old, completed the MSSQ and the CSQ-VR, and were immersed in VR. During immersion, they were exposed to a roller coaster ride. Before and after the ride, participants responded to CSQ-VR and performed VR-based cognitive and psychomotor tasks. Post VR session, participants completed the CSQ-VR again. Results: Motion sickness susceptibility, during adulthood, was the most prominent predictor of cybersickness. Pupil dilation emerged as a significant predictor of cybersickness. Experience in videogaming was a significant predictor of both cybersickness and cognitive/motor functions. Cybersickness negatively affected visuospatial working memory and psychomotor skills. Overall cybersickness', nausea and vestibular symptoms' intensities significantly decreased after removing the VR headset. Conclusions: In order of importance, motion sickness susceptibility and gaming experience are significant predictors of cybersickness. Pupil dilation appears as a cybersickness' biomarker. Cybersickness negatively affects visuospatial working memory and psychomotor skills. Cybersickness and its effects on performance should be examined during and not after immersion.

## 1. Introduction

Immersive Virtual Reality (VR) represents one of the most remarkable technological advancements of the 21st century. A digital interface that promises full immersion into an alternate or simulated environment, VR's potential has been tapped across an array of disciplines. Entertainment industries have been early adopters, providing audiences with experiences that were once relegated to the realms of imagination [1]–[3]. Concurrently, the education sector has witnessed a paradigm shift with VR-enhanced pedagogical tools, fostering enriched learning experiences [4]–[6].

Furthermore, the domain of professional training has embraced VR to craft realistic scenarios for a myriad of professionals, from manual labourers mastering their craft to surgeons simulating complex procedures [7]–[9]. The medical arena is reaping the benefits too. Beyond conventional treatments and therapies, VR is emerging as a powerful adjunctive tool. Pain management, once reliant solely on pharmacological interventions, now explores the pain-distracting potential of VR [10]. Rehabilitation, be it physical or neurological, is experiencing a renaissance with VR-infused therapies [11], [12].

Neuropsychology, in particular, has found a robust partner in VR, aiding in cognitive assessments [13], [14], training [15], and targeted rehabilitation efforts [16]–[18].

Such broad applications further extend their arms to vulnerable populations. The elderly, often considered tech-averse, find solace and cognitive rejuvenation in VR experiences tailored to their needs [19], [20]. Individuals with Mild Cognitive Impairment (MCI) and/or a type of dementia [21], [22], or developmental challenges such as Attention-Deficit Hyperactivity Disorder (ADHD) [23]–[25]) and Autism Spectrum Disorder (ASD) [17], [26], [27] are not mere spectators. VR interventions, designed with sensitivity and precision, are being administered to provide these populations with therapeutic as well as recreational relief. Considering the importance of the aforementioned applications of VR and the fragility of some of the targeted populations, the effective implementation of VR becomes imperative.

### 1.1. Cybersickness in Virtual Reality

While the promise of VR in transforming various domains is indisputable, VR also harbours an inherent limitation—cybersickness, a condition affecting a segment of its users [28]. Manifesting as a triad of nausea, disorientation, and oculomotor disturbances, cybersickness remains a significant concern. While there's a temptation to draw parallels between cybersickness and simulator sickness, the two display distinct characteristics [29]. Notably, cybersickness presents with heightened general discomfort, particularly intensified by nausea and disorientation [29]. Adding another layer to this complex tapestry, cybersickness also stands apart from motion sickness. The former arises primarily from visual cues in VR, whereas the latter emerges from actual movement [30].

Delving into the underlying causes of cybersickness, a comprehensive theoretical understanding is still emerging. However, the sensory conflict theory has gained traction, shedding light on the root of the issue [28], [30], [31]. According to this theory, the unsettling symptoms of cybersickness predominantly result from a sensorial conflict between the visual system and the vestibular mechanisms in our inner ear [28], [31]. At its core, our sense of balance and spatial positioning is an intricate intertwining of visual, vestibular, and proprioceptive feedback. These systems often receive mis-matched cues in VR, leading to sensory dissonance. For VR, this conflict can be at-tributed to vection—an illusionary perception of motion [32]. This illusion, particularly when accompanied by movements like linear and angular accelerations, has been pinpointed as a major instigator of cybersickness in VR [33], [34]. As VR continues its ascendancy in the tech world, the quest to understand and alleviate cybersickness remains a pressing concern.

### 1.2. Mitigation of Cybersickness

Two primary catalysts for cybersickness are VR's hardware and software characteristics [35]. Hardware issues such as latency, which refers to the delay between a user's action and the corresponding change in the virtual environment, can be incredibly disorienting. Similarly, refresh rates that are not synchronized with a user's natural perception can lead to a jarring experience [35]. On the software side, inconsistencies in visual-vestibular integration, where what one sees doesn't match what one feels, can throw off the body's equilibrium [35], [36]. Similarly, increased cognitive workload and confusion may also play a role in modulating cybersickness intensity [35], [36].

However, the industry and the scientific community are not passive in the face of these challenges. With advancements in head-mounted displays (HMDs), many of these hardware-related issues are gradually being addressed [35]. Better display resolutions, faster refresh rates, and improved motion tracking reduce the disconnect users feel. Simultaneously, on the software end, industry or research software developers are now more attuned to creating experiences that align with human physiology. Guidelines tailored for specific scientific fields and targeted populations are emerging, ensuring a more holistic and comfortable VR experience for all [36]–[41].

Furthermore, various strategies have been employed to counteract the symptoms of cybersickness, each with its own set of challenges. One such strategy is "acclimatization", which involves frequently exposing a person to elements that induce cybersickness to help them build resilience against it [42]. While this might be an effective solution, it is both time-intensive and costly, demanding significant effort and dedication. Additionally, there are recommendations for using specific medications or natural remedies, such as ginger, to tackle the effects [43], [44]. However, these methods can be obtrusive and might introduce unintended consequences like fatigue or potential allergies [43].

In the realm of Human-Computer Interaction (HCI), innovations such as adjusting the user's perspective [45], introducing dynamic focus shifts [46], ensuring bodily equilibrium [47], and incorporating brief intervals [48] have been put forward. However, these methods can impede certain virtual interactions (e.g., adjusting user perspective and maintaining balance) and might detract from the overall immersive experience (e.g., focus shifts and intervals). Similarly, joyful and calming music effectively alleviates cybersickness symptoms in VR [49]. However, again, using music is only suitable for some virtual environments, given that may confound the purpose of this VR application. An efficient approach, which may be a universal solution, is to detect and prevent cybersickness. However, this universal solution should be adaptive to the user's needs and requirements.

### 1.3. Questionnaires and Physiological Metrics of Cybersickness

One of the most common approaches to measure cybersickness is the administration of questionnaires, typically before and after exposure to VR [36], [50]. The effectiveness of these tools in evaluating cybersickness in VR settings has been a topic of keen interest. The Simulator Sickness Questionnaire (SSQ) [51] has received criticism from several studies for its inability to adequately measure cybersickness in VR [52]–[55]. While the Virtual Reality Sickness Questionnaire (VRSQ) was conceived as an improvement on the SSQ, it too exhibits considerable limitations, especially in terms of its structure and specificity [53], [56].

On the other hand, the Cybersickness in Virtual Reality Questionnaire (CSQ-VR) emerges as a more reliable and comprehensive tool [49], [53], [57], [58]. Given that intense cybersickness can significantly hamper a user's cognitive and motor functions, especially in reaction times [34], [49], [53], [59], [60], it's imperative for a tool to detect these declines effectively. The CSQ-VR, with its robust design and metrics, compared to both SSQ and VRSQ, was substantially more efficient in detecting cognitive and/or motor skills decline due to cybersickness. Furthermore, the CSQ-VR is offered in a 3D-VR version, allowing for a repeated evaluation of cybersickness while the user is immersed in VR. This is essential considering that the intensity of cybersickness can vary throughout a VR session [32], [49], [53], [61]. Additionally, considering that pupil size was a significant predictor of cybersickness' intensity [49], [53] the integration of eye-tracking in the CSQ-VR further augments its cybersickness' assessment capabilities.

Beyond questionnaires, and the pupil size discussed above, other physiological objective metrics have been used to examine and predict cybersickness symptomatology [30], [62]. In this direction, electroencephalography (EEG), which captures electrical activity in the brain's cortical areas, is an efficient approach to detect cybersickness [63]. However, using an EEG in combination with the VR HMD is not feasible for widespread utilization due to the high cost of an EEG, as well as that is ergonomically problematic having both cumbersome devices mounted on the head of the user. Other physiological metrics encompass electrocardiogram (ECG), electrogastrogram (EGG), electrooculogram (EOG), photoplethysmogram via pulse oximeter (PPG), breathing rate, and galvanic skin response (GSR), which have been found to predict cybersickness symptomatology and intensity [62], [64]. For example, rises in bradygastric activity, breathing rate [64], [65], heart rate [60], and forehead skin conductance [66], [67] offer reliable indicators of cybersickness. However, beyond PPG and CSR, the other devices are also costly and not ergonomically appropriate for effective use. On the other hand, PPG and GSR can be embedded in haptic gloves (e.g., see TeslaGloves) and be ergonomically efficient [68], [69]. Nevertheless, these haptic devices are costly, so, beyond applications in industry or enterprises, these devices are not affordable for the general population [68], [69]. Finally, pupil size is not the only eye-tracking metric that may indicate cybersickness. Other eye-tracking metrics, such as fixation duration and distance between the eye gaze and the targeted object, may assist in predicting cybersickness [70]. Therefore, except for the eye-tracking that is embedded in many VR HMDs, the use of the rest of the physiological metrics is either ergonomically and/or financially problematic.

### 1.4. Individual Differences and Cybersickness

The experience of cybersickness varies among individuals, with factors like gender playing a potential role [49], [50]. Some studies suggest female users may experience more intense cybersickness than male users, but findings have been inconsistent [49], [50]. For instance, Petri et al.'s study [71] found no significant gender differences in objective metrics like heart rate but did observe a difference in subjective experiences based on the SSQ. Meanwhile, Melo et al. [72] found no such gender differences. Similarly, Stanney et al.'s experiments [73], [74] found gender insignificant in predicting cybersickness in certain conditions. A meta-analysis also supported the absence of significant gender differences when evaluating cybersickness in VR settings using the SSQ [50]. However, it's speculated that gaming and VR experience might play a role.

Few studies have delved into the impact of computing, VR, or gaming experience on cybersickness. While Stanney et al. [73] considered gaming experience, their methodology regarding its measurement was ambiguous, and their results, postulating an absent effect, cannot thus be considered reliable. Kourtesis and collaborators in a series of studies [13], [75], [76] found no significant effect of gaming or VR experience on cybersickness. However, the VR software implemented in these studies was thoroughly designed and developed to elicit minimal to no cybersickness symptoms. Conversely, Weech et al. [77] found that gaming experience might influence cybersickness, especially when paired with narratives. Likewise, in the recent study of Kourtesis et al. [49] gaming experience was found to affect experiencing cybersickness, where higher gaming experiences indicated a higher resilience. Also, the same study showed that gaming experience explained gender differences in terms of cybersickness, where participants of the opposite sex but with the same gaming experience did not experience different cybersickness' intensity [49].

Moreover, each individual may demonstrate a diverse level of susceptibility to experiencing cybersickness [78], [79]. Previous studies, using the Motion Sickness Susceptibility Questionnaire (MSSQ) scores, which measure susceptibility to experiencing intense symptoms of motion sickness [80], were found to be associated with personality traits and anxiety levels [81], [82]. Similar to motion sickness susceptibility, visually induced motion sickness (i.e., cybersickness induced by vection) demonstrates similar patterns of susceptibility among individuals [78], [79]. Taking these together, given the similarities between motion sickness and cybersickness elicited by vection, the MSSQ scores may indicate susceptibility to cybersickness. However, in previous studies, the MSSQ did not predict cybersickness intensity [49], [53]. Nevertheless, the aforementioned studies also used MSSQ to exclude participants with moderate-to-high motion sickness susceptibility. Thus, MSSQ's utility in predicting cybersickness has not yet been examined appropriately. Overall, while individual differences influence cybersickness, the exact factors and their interplay remain a complex topic of investigation.

### 1.5. Effects of Cybersickness on Cognitive and Motor Skills

Apart from impacting the quality of the user experience in VR, cybersickness can also detrimentally influence a user's cognitive and motor functions. Considering VR's applications in areas demanding unimpaired cognitive and motor skills, like education, research, clinical settings, and training, cybersickness poses significant challenges to VR's successful integration. Several systematic reviews [35], [50], [83] have highlighted a considerable, though temporary, decline in cognitive and/or motor functions due to cybersickness in immersive VR. Dahlman et al. [84] theorized that motion sickness notably impairs users' verbal working memory. Similarly, a study by Varmaghani et al. [85] with 47 participants split into VR and control groups found that the VR group did not experience the expected enhancement in visuospatial skills seen in the control group, suggesting cybersickness's potential effect on visuospatial learning.

Mittelstaedt et al.'s research [59] explored cybersickness's influence on various cognitive areas before and after VR exposure. The study revealed that cybersickness resulted in delayed reaction speeds and hindered the anticipated boost in visual processing speed, indicating a negative impact on attention and reaction time. However, spatial and visuospatial memory skills seemed unaffected. Similarly, studies by Nalivaiko et al. [60] and Nesbitt et al. [34] reported slowed reaction times correlated with increasing cybersickness severity, suggesting a link between cybersickness intensity and potential cognitive/motor deterioration. It's important to note that while these studies underscore cybersickness's negative effects on cognitive and/or motor abilities, they assessed cybersickness post-VR exposure, not during the experience itself.

Since the users experience a readjustment to physical space, while removing the VR HMD and transitioning from the virtual to the physical environment [86], examining cybersickness, cognition, and motor skills after exposure may confound the observations. . In a recent study examining cybersickness, cognitive and motor skills during immersion, cybersickness had a significant negative effect on verbal working memory and psychomotor skills, but not on visuospatial working memory [49]. However, the order of the tasks was not randomized and counterbalanced, and the researchers, using the MSSQ, excluded participants who showed high susceptibility to experiencing motion sickness. Thus, based on the studies mentioned above, while evidence postulates a negative impact of cybersickness on cognitive and motor skills, there are still discrepancies, especially regarding the size of these effects.

### 1.6. Current Study Aims

Based on the review of existing literature, it is clear that VR is a cutting-edge tool with extensive applications spanning from entertainment to research and rehabilitation. However, the applications of VR may be hindered by the onset of cybersickness. There are discrepancies about the impact of cybersickness on core cognitive functions, which are required in VR applications in several fields (e.g., education, occupational training, and rehabilitation). Also, the role of individual differences (e.g., IT skills, sex, and susceptibility to cybersickness) in experiencing cybersickness has not been fully understood. Given this background, the research inquiries have been articulated in the following hypotheses:

H1: Pupil size will be a significant predictor of the intensity of cybersickness.

H2: Susceptibility to motion sickness will be a significant predictor of the intensity of cybersickness.

H3: Computer and/or videogame experience will predict the intensity of cybersickness symptomatology.

H4: Cybersickness symptomatology will have a negative effect on verbal working memory, visuospatial working memory, and/or psychomotor skills.

H5: The cybersickness intensity during and after VR exposure will be significantly different.

## 2. Materials and Methods

### 2.1. Virtual reality Hardware & Software

An HTC Vive Pro Eye was used, which embeds an eye-tracker with a binocular gaze data output frequency of 120Hz (i.e., refresh rate), a tracking accuracy of 0.5°-1.1°, a 5-point calibration, and a 110° trackable field of view. The HTC Vive Pro Eye substantially surpasses the minimum hardware criteria for alleviating and/or avoiding cybersickness [35]. Thus, beyond facilitating the collection of eye-tracking metrics, its utilization further ensures that the linear and angular accelerations will induce cybersickness (see description be-low) in the virtual environment and not by hardware inadequacies. Likewise, the software development was performed in line with the guidelines for VR software for research and clinical settings, which have been found efficient in substantially mitigating cybersickness symptomatology [14], [39]. This further ensured the avoidance or decrease of effects of software characteristics on the expression or intensity of cybersickness.

### 2.2. Virtual Environment Development

The virtual environment was based on the one used in our previous studies on cybersickness [49], [53]. The Unity3D game engine was utilized to develop the virtual environment. Also, the SteamVR SDK aided in the creation of interactions. Given the potential influence of gaming experience on task performance [14], the virtual hands/gloves feature of the SteamVR SDK was integrated to enable straightforward interactions. Crucially, these interactions were designed to be intuitive, initiated by touching the object for selection and maintained touch for confirmation, eliminating the need for button presses. The virtual gloves offered by SteamVR were neutral, not hinting at any particular gender or race, which helped mitigate any biases tied to these factors [87].

For audio instructions, Amazon Polly was used to generate neutral, realistic voice clips. To ensure clarity and user comprehension, instructions were offered in video, audio, and text formats, promoting smooth task execution. The SteamAudio plugin was employed for spatial audio effects, particularly feedback sounds. Eye-tracking and pupillometry, including monitoring pupil size, fixation counts, fixation duration, and distance of the object from the eye, were facilitated through the SRapinal SDK. The bmlTUX SDK [88] enabled the randomization of experimental sequences, data exportation into a CSV format, and overall experimental management.

### 2.3. Roller Coaster Ride: Linear and Angular Accelerations

The design of the roller coaster ride in this research was modelled after the rides used in our prior cybersickness studies [49], [53]. A 12-minute ride was designed, through which each participant had to undergo to experience linear and angular accelerations throughout the ride. The trajectory was animated to represent the moving platform the user stood on (see Figure 1). The general direction of movement was forward, with an exception towards the end (refer to reverse z-axis). Overall, the platform's motions mimicked the dynamics of a roller coaster.

The designed route encompassed a specific sequence of accelerations: (1) linear on the z-axis, (2) angular involving z- and y-axes, (3) angular encompassing z-, x-, and y-axes, (4) angular on the roll axis, (5) heightened linear acceleration on the z-axis, (6) angular on the yaw axis, and (7) intense linear acceleration involving y-axis followed by an inverse on the z-axis. The environment was kept minimalistic, predominantly in shades of black and white (as illustrated in Figure 1). This design was selected to minimize extraneous factors that could influence or elicit cybersickness symptoms. Moreover, the tiled pattern provided visual references, helping users discern changes in direction and altitude.

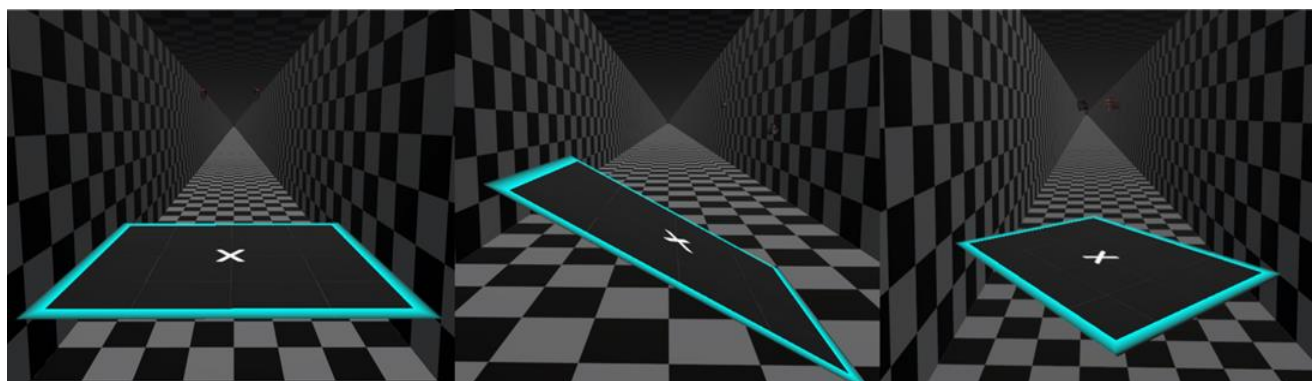

Figure 1. Examples of Linear and Angular Accelerations During the Ride.

### 2.4. Cognitive & Motor Skills Assessment in Virtual Reality

Due to the need for repeated evaluations of cybersickness, cognition, and motor abilities, while users are immersed in VR, immersive VR iterations of recognized tasks/tests were used (see Figure 2). These VR cognitive and motor skills tests have been used in our previous studies [49], [53]. In designing these VR-based cognitive and psychomotor tasks, we adhered to the specific design principles and development guidelines for cognitive evaluations in immersive VR as outlined in [14], [39]. Also, given that these tasks necessitate physical actions, their design was aligned with the ISO 9241-400:2007 standards from the International Organization for Standardization, which focus on the ergonomics of human-system interaction ([89]. This approach, which includes considerations like personalizing object height and the distance between the shoulder and the object, has been used in a prior study [90].

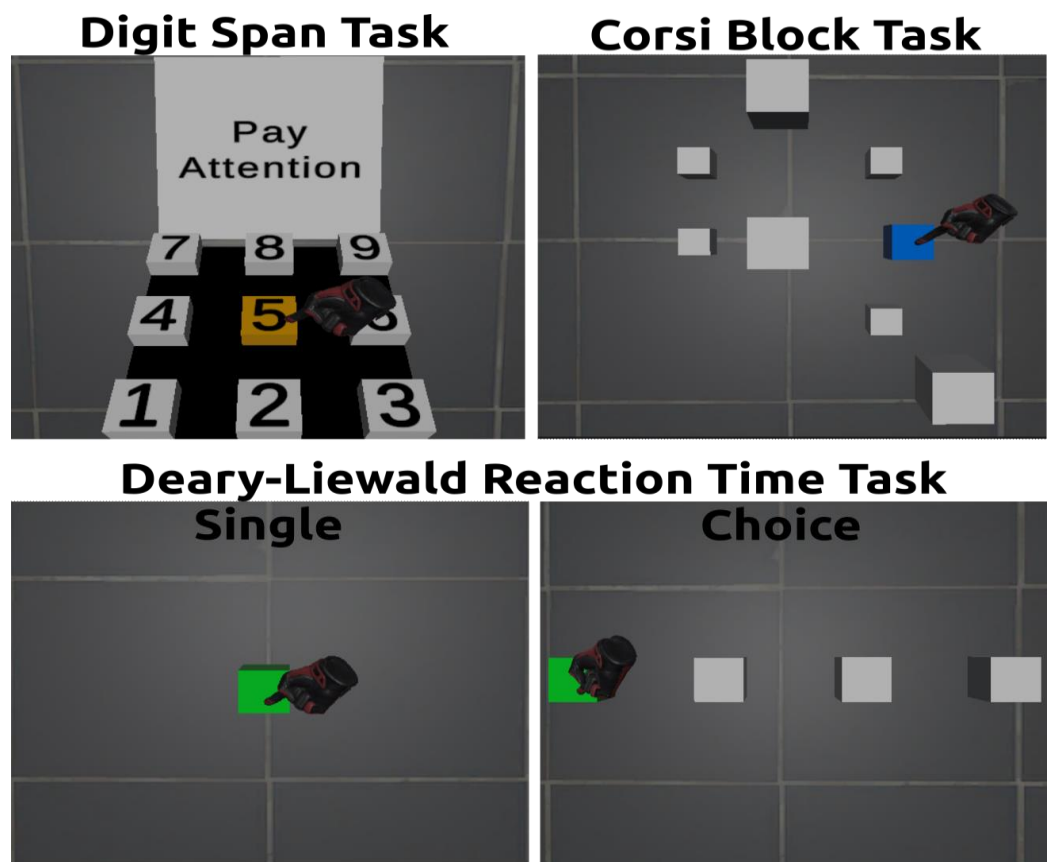

Figure 2. Digit Span Test (Upper Left), Corsi Block Test (Upper Right), and Deary-Liewald Reaction Time Test (Bottom).

### 2.4.1. Verbal Short/Working Memory: Digit Span Test.

A VR version of the Digit Span Test (DST) [91] was developed and used (see Figure 2). The test encompasses two tasks, the forward and backward recall tasks, which examine verbal short-term memory and verbal working memory respectively [92]. The order of administration of the tasks is standardized, where examinees perform first the forward recall task, and last the backward recall task [91], [92]. In the DST, participants were presented with a sequence of numbers to listen to, which they were then required to recall in the same (i.e., forward recall task) and reverse order (i.e., backward recall task). For instance, if the given sequence was 2, 4, 3, to correspondingly respond with 2, 4, 3 (i.e., forward recall task); or 3, 4, 2 (i.e., back-ward recall task). Post-listening, a virtual keypad materialized for the participants to input their answers. To select a digit, they touched the corresponding white box on the keypad (see Figure 2). This box turned blue upon touch. A continuous touch for more than a second provided confirmation of their selection. Correct selections turned the button green accompanied by an affirmative sound, while incorrect ones made it turn orange with a disapproving sound. A trial concluded either upon an error or once the sequence was correctly inputted. With every alternate correct trial, the sequence length increased. The task terminated after two consecutive errors at the same sequence length (e.g., three digits) or after completing the maximum sequence length (two trials of 7 digits). The Total Score combined the count of correctly completed trials and the longest digit sequence (i.e., digit span). A video description of the test can be accessed via the following link: https://www.youtube.com/watch?v=1H8cqci-lFs .

### 2.4.2. Visuospatial Short and Working Memory: Corsi Block Test.

The VR version of the Corsi Block Test (CBT) [93] was developed and employed (see Figure 2). Comparable to the DST described above, the CBT incorporates the forward recall task and the backward recall task, which evaluate visuospatial short-term memory and visuospatial working memory correspondingly [93], [94]. The sequence in which tasks are performed is also standardized: participants first perform the forward recall task and then conclude with the backward recall task [93], [94]. Each task displays 27 white cubes, positioned distinctively across the x, y, and z dimensions. However, nine of these 27 cubes were randomly showcased to participants in each trial (see Figure 2). Each trial began with nine cubes. Based on the current sequence length, a subset of these cubes would illuminate in blue for a second each, accompanied by a bell sound. Once this sequence finished, participants needed to recall and select the cubes in the same (i.e., forward recall task) or reverse order (i.e., backward recall task). A cube was selected by touching it and turning it blue. Maintaining the touch for a second confirmed the choice: the cube turned green and sounded a positive tone if correct, or orange with a negative tone if wrong. The trial ended upon a mistake or once the full normal (i.e., forward recall task) or reversed (i.e., backward recall task) sequence was recalled correctly. Trials started with sequences of two cubes, increasing by one if at least one of the two attempts was correct. The task concluded after two errors at a specific sequence length or reaching the maximum sequence of seven cubes. Perfect performance meant achieving up to seven cubes without errors. The Total Score was the sum of the highest achieved sequence length (i.e., Corsi block span) and the total number of correctly recalled sequences. A video description of the test can be accessed via the following link: https://www.youtube.com/watch?v=MLiIvkyMt-g .

### 2.4.3. Psychomotor Skills: Deary-Liewald Reaction Time Test.

The VR version of the Deary-Liewald Reaction Time test (DLRT) [95] test was developed and used to evaluate psychomotor skills. The DLRT incorporates two tasks: the simple reaction time (SRT) task and the choice reaction time (CRT) task [95]. In the SRT task, participants watched a white box, which they had to quickly touch whenever it turned blue (see Figure 2). This was repeated for 20 trials. For the CRT task, any one of four horizontally aligned boxes would randomly turn blue, prompting participants to touch it as fast as they could (see Figure 2). This occurred over 40 trials. In both tasks, participants were directed to touch the highlighted boxes using either hand swiftly. Before starting, a practice round ensured that participants grasped the instructions. The SRT score was derived by averaging reaction times across its 20 trials, and similarly for CRT. However, CRT provided three separate scores. Using eye-tracking, we gauged the time to notice the target (i.e., Attentional Time) and the time between noticing and touching the target (i.e., Motor Time). An overall reaction time, from target appearance to touch, was also recorded. Thus, three scores emerged:

1) Reaction Time (RT) reflecting overall psychomotor speed.

2) Attentional Time (AT) showing attention processing speed.

3) Motor Time (MT) representing movement speed.

A video description of the test can be accessed via the following link: https://www.youtube.com/watch?v=wXdrt0PjNsk .

### 2.5. Cybersickness in Virtual Reality Questionnaire (CSQ-VR)

To assess the symptoms and intensity of cybersickness, the CSQ-VR was used, which is a valid tool for assessing cybersickness and has shown superior psychometric properties to the SSQ and the VRSQ [53]. Also, originating from the VR-Induced Symptoms and Effects section of the VR Neuroscience Questionnaire, the CSQ-VR boasts strong structural and construct validity [36]. Its strengths include a concise format (just six questions) and the generation of comprehensible results [53], [58]. Additionally, it captures various cybersickness subcategories, such as nausea, disorientation, and oculomotor disturbances. Each category has two questions, scaled on a 7-point Likert Scale, with options ranging from "1 - absent feeling" to "7 - extreme feeling" — each option combines text and a number (see Figure 3). From the CSQ-VR, a total score and three subscores (i.e., one for each subcategory: Nausea, Disorientation, and Oculomotor) can be derived. The total score is the summation of the three subscores. The paper-and-pencil version of the CSQ-VR was administered twice, before and after exposure to VR.

Given the study's goal of repeatedly measuring cybersickness during VR immersion, the 3D-VR version of CSQ-VR was implemented. The question appeared at the top in the designed user interface, with the chosen response (in red) situated centrally. Users could modify their answers using a slider, by selecting a number directly or sliding along the scale (see Figure 3). The VR version of CSQ-VR also incorporates eye-tracking metrics. Invisible tracking markers were positioned ahead of the text, and their dimensions always matched the visible text per line (see Figure 3). This setup allow us to gauge the Fixation Duration over the text as an indicator of reading rate. Additionally, continuous pupil measurement occurred while users engaged with the CSQ-VR, allowing us to determine the average pupil size (for both eyes) during interactions, which has been previously seen as a biomarker of cybersickness intensity [49], [53]. A video description of the 3D-VR version of CSQ-VR can be accessed via the following link: https://www.youtube.com/watch?v=npW4NKNLXok.

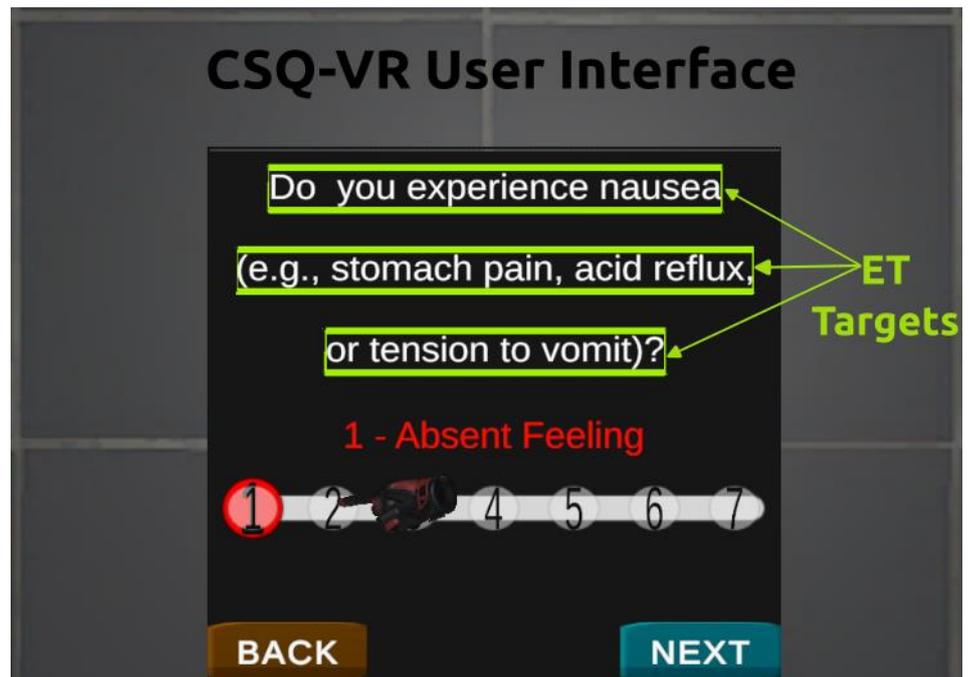

Figure 3. CSQ-VR User Interface & Eye-Tracking (ET) Targets. Note: eye-tracking targets were not visible to the user.

### 2.6. Demographics and Motion Sickness Susceptibility Questionnaire (MSSQ)

A custom questionnaire was used to collect demographic information, such as gender, age, education, and computer and videogame skills. Smartphone/computing/gaming experiences were determined by summing up scores from two specific questions, each based on a 6-point Likert scale. The initial question gauged the participant's proficiency or skill level in using smartphone/computer/games, with ratings like '5: highly skilled.' In contrast, the subsequent question focused on how often they engage with these platforms, with responses such as '4: once a week.' This custom questionnaire is the same that has been used in previous studies (e.g., [49], [53], [90]). Also, the short form of the Motion Sickness Susceptibility Questionnaire (MSSQ) was used to measure participants' predisposition to experience motion sickness [80]. MMSQ serves as a diagnostic instrument designed to gauge an individual's vulnerability to motion sickness and pinpoint specific triggers associated with the onset of the symptoms. It delineates experiences into two distinct categories:

1. <u>Childhood Experience (prior to the age of 12)</u>: Here, respondents specify the frequency with which they encountered sensations of sickness or nausea in different modes of transport or specific entertainment scenarios.
2. <u>Experience over the Last 10 Years</u>: This section requires individuals to recount the number of times they felt symptoms of sickness or nausea under similar circumstances but within the past decade.

Each section receives an independent score, and the cumulative result from both areas offers the raw MSSQ score. This raw score can be translated into a percentile through reference tables or a dedicated polynomial for enhanced interpretability. Thus, three scores were derived from MSSQ: the MSA-Child, the MSB-Adult, and the MSSQ-Total. However, it is pivotal to comprehend that the term "sickness" within the MSSQ's framework encapsulates feelings ranging from mere queasiness to outright nausea or even vomiting. Notably, the MSSQ has clinical significance as it can indicate inherent motion sickness susceptibility, especially valuable for those diagnosed with vestibular diseases, as well as examining susceptibility to visually induced cybersickness in VR [78]–[82].

### 2.7. Participants & Experimental Procedures

Participants were recruited through convenience sampling, utilizing the internal mailing lists of the National and Kapodistrian University of Athens, as well as promotions on social media platforms. The research received approval from the Ethics Committee of the Department of Psychology of the National and Kapodistrian University of Athens. The sample consisted of 30 participants, 17 women and 13 men, aged 20 to 45 years old. All participants had normal or corrected vision using contact lenses or glasses. Upon arrival, all participants were informed about the procedure and gave their informed consent in written form before proceeding with the experimental procedures. The process began with the completion of the demographic data questionnaire and the questions of the MSSQ questionnaire.

Then, before using the VR, the participants were asked to complete the CSQ-VR questionnaire in a paper-and-pencil form. Only then did participants go through an induction to VR technology and how to use and wear the HTC Vive Pro Eye. The in-duction also described the cognitive and motor tasks they had to perform in VR. After the induction, under the guidance of the experimenter, the participants wore the VR HMD and performed the eye-tracking calibration as offered by the HTC Vive in collaboration with the SteamVR. Once in VR, they stood at a designated spot marked 'X' (see Figure 1). The session began with tutorials. For each task, a video tutorial, along with verbal and written guidelines, was provided.

During immersion, the first task participants had to perform was to provide their responses to the questions of the 3D-VR version of the CSQ-VR. After that, the participants had to perform the VR versions of the cognitive (i.e., DST and CBT) and psychomotor skills (i.e., DLRT) tests (see Figure 2). After going through the tutorial of each task, where participants received audiovisual (i.e., audio, video, and labels) information about the task, they then performed the corresponding task. This stage of going through the tutorials and performing all the tasks served as the baseline assessment, which lasted approximately 25 minutes. Note that the order of the tests was counterbalanced across participants. Given that the tests were DST, CBT, and DLRT, three different orders were used:

1. DST, CBT, and DLRT.
2. CBT, DLRT and DST.
3. DLRT, DST, and CBT.

Thus, the full circle of orders was complete for every three participants. In the whole sample, the circle of orders was completed ten times (i.e., 30 participants divided by three orders).

The roller coaster ride followed, which lasted about 12 minutes after this baseline assessment. After the ride, participants performed the same set of assessments as the baseline (i.e., CSQ-VR, DST, CBT, and DLRT). Note that the order of administration of the tests was identical to the baseline assessment. Participants thus underwent one ride of 12 mins and two assessment sessions (i.e., before and after the ride). The entire VR procedure spanned roughly 60 minutes for every participant. Finally, after removing the VR HMD (i.e., post-VR exposure), participants completed the paper-and-pencil version of the CSQ-VR questionnaire. Then, refreshments rich in electrolytes were offered to the participants, and they were given a 10–15-minute rest before departure. Before leaving, participants were advised against driving and operating heavy machinery for the rest of the day.

### 2.8. Statistical Analyses

All the statistical analyses were conducted using R language [96] in RStudio [97]. Furthermore, the psych (correlational analyses & t-tests) [98], the ggplot2 (plots) [99], and the lme4 (regression analyses) [100] R packages were used to perform the respective analyses. Descriptive statistical analysis was performed to provide an overall review of the sample. The paired samples t-test examined the differences in intensity of cybersickness between during and after exposure to VR. Finally, multiple linear regression analyses were performed to examine the predictors of cybersickness symptomatology, and multiple mixed linear regression model analyses were performed to examine the predictors of performance on cognitive and psychomotor skills' tasks. As the variables violated the condition of normality, the bestNormalize R package [101] was used to transform (e.g., into logarithms) and center the data (i.e., converting them into z scores). In this way, there was a normal distribution in the data to conduct parametric statistical analyses..

2.4.1 Regression Analyses.

The assumption of normality for event- and time-based scores was evaluated using the Shapiro–Wilk Normality test. The Non-Constant Error Variance test was applied to verify the homoscedasticity assumption for the models. Multicollinearity was assessed by calculating the variance inflation factor for the predictors in each model. Linear regression analyses were utilized to explore the predictors of everyday PM functioning based on everyday cognitive functions. Model comparisons were made using analyses of variance. The comparison criteria among the models included Akaike's information criterion (AIC), Bayesian criterion (BIC), the F statistic and its significance level, and the proportion of variance explained (i.e., $R^2$).

In the analytical approach adopted, a wide array of variables was considered as potential predictors for the models. Specifically,

For the Linear Multiple Regression analyses, the following variables were considered: such as Sex, Education, Age, Computing Experience, Smartphone Apps Experience, Gaming Experience, MSA-Child, MSB-Adult, MSSQ Total, Pupil Size during responding to CSQ-VR, and Pupil Size during the Ride.

For the Mixed Model Regression analyses (i.e., prediction of performance on cognitive and psychomotor skills' tasks) the following variables were considered: CSQ-VR Total Score, CSQ-VR Nausea Score, CSQ-VR Vestibular Score, CSQ-VR Oculomotor Score, Sex, Education, Age, Computing Experience, Smartphone Apps Experience, and Gaming Experience.

With these variables as the foundation, a systematic and incremental model development process was initiated:

Single-Predictor Models: Initially, separate models were developed. In each of these models, a single cognitive predictor from the list was incorporated. When the performance of these models was compared, the individual variable that yielded the best preliminary results was identified.

Dyadic Predictor Models: A new set of models was crafted in the subsequent phase, each containing two predictors. The top-performing variable from the first step was consistently used as one of the predictors in these models. The second predictor was drawn sequentially from the remaining list of variables. After the two-predictor models had been designed, their performances were critically evaluated. The top-performing two-predictor model was then juxtaposed with the best model from the first step.

Subsequent Iterations in the Incremental Approach: The methodology was characterized by its iterative nature. In each subsequent phase, the best predictors from the previous step were retained, and an additional predictor from the list was introduced. With each iteration, the models were rendered more complex. This step-by-step comparison continued until a point was reached where the introduction of more variables failed to enhance the model's performance significantly. When a model from an earlier step outperformed a more complex model from a subsequent step, it indicated that the optimal combination of predictors had been identified, and the simpler model from the previous step was chosen as the final best model. This rigorous process ensured that the final model was robust and represented the best combination of the variables initially considered.

## 3. **Results**

Descriptive statistics of the data are presented in Table 1. Demographically, participants were relatively young adults with a wide range of educational backgrounds. Their technology engagement was evident from the experience they had with computing, smartphones, and gaming. This provides an interesting lens through which to understand the effects of VR, given their wide range of familiarity with digital tools. Regarding motion sickness, participants displayed varied susceptibility. Their scores from childhood to adulthood in motion sickness susceptibility showed a notable shift, suggesting that reactions to motion may evolve with age. When diving into the cybersickness, as assessed by the CSQ-VR, it seemed to intensify post-VR exposure across all its subcategories: nausea, vestibular, and oculomotor symptoms.

**Table 1.** Descriptive statistics

|  | *Stage* | *Mean (SD)* | *Range* |
|---|---|---|---|
| Age | – | 29.17 (4.17) | 22-38 |
| Education | – | 16.27 (2.92) | 12-25 |
| Computing XP | – | 10.30 (1.05) | 8-12 |
| Smartphone XP | – | 10.70 (0.75) | 9-12 |
| Gaming XP | – | 37.23 (14.64) | 22-7 |
| MSA – Child | – | 6.02 (4.38) | 0-14.14 |
| MSB – Adult | – | 3.61 (3.18) | 0- 12.60 |
| MSSQ Total | – | 9.63 (6.73) | 0-26.74 |
| CSQ-VR – Total | Baseline | 8.73 (3.06) | 6-19 |
| | Post-Ride | 14.78 (7.84) | 6-34 |
| CSQ-VR – Nausea | Baseline | 2.50 (0.90) | 2-6 |
| | Post-Ride | 5.20 (3.22) | 2-14 |
| CSQ-VR – Vestibular | Baseline | 2.63 (0.89) | 2-5 |
| | Post-Ride | 4.63 (3.18) | 2-14 |
| CSQ-VR – Oculomotor | Baseline | 3.60 (1.70) | 2-8 |
| | Post-Ride | 4.93 (2.87) | 2-12 |
| DST – Forward | Baseline | 16.63 (2.58) | 12-20 |
| | Post-Ride | 16.90 (2.63) | 11-20 |
| CBT – Forward | Baseline | 14.63 (2.72) | 10-20 |
| | Post-Ride | 15.17 (3.05) | 5-20 |
| DST – Backward | Baseline | 14.83 (3.58) | 8-20 |
| | Post-Ride | 15.63 (3.64) | 7-20 |
| CBT – Backward | Baseline | 15.10 (2.00) | 11-19 |
| | Post-Ride | 14.43 (2.89) | 8-20 |
| DLRT - RT | Baseline | 0.57 (0.10) | 0.36-0.81 |
| | Post-Ride | 0.61 (0.16) | 0.38-1.13 |
| DLRT - AT | Baseline | 0.30 (0.09) | 0-0.42 |
| | Post-Ride | 0.34 (1.44) | 0- 0.87 |
| DLRT - MT | Baseline | 0.24 (0.10) | 0-0.45 |
| | Post-Ride | 0.27 (0.17) | 0- 0.83 |

XP = Experience; MS = Motion Sickness; MSSQ = Motion Sickness Susceptibility Questionnaire; CSQ-VR = Cybersickness in Virtual reality Questionnaire; DST = Digit Span Test; CBT = Corsi Block Test; DLRT = Deary-Liewald Reaction Time Test; RT = Reaction Time; AT = Attention Time; MT = Motor Time;

### 3.1. Linear Regression Analyses: Prediction of Cybersickness Intensity

Linear regression analyses were conducted to detect and compare the significant predictors of overall, per symptom category, cybersickness intensity. Table 2 elucidates how various individual factors predict overall cybersickness. In support of **H1**, the pupil size during responding to the CSQ-VR questionnaire and the VR ride demonstrated significant cybersickness prediction. In agreement with **H2**, the most significant predictor is the Motion Sickness susceptibility score as an adult (MSB-Adult), with an impressive 39% variance ($R^2=0.39$) accounted for. Finally, in line with **H3**, smartphone experience and gaming experience were notable for their significance in predicting cybersickness.

**Table 2.** Single Predictor Models for Overall Cybersickness

| Predictor | β coefficient | p-value | $R^2$ |
|---|---|---|---|
| Sex (Male) | -0.71& | .053 | 0.13 |
| Education | 0.14 | .470 | 0.02 |
| Age | -0.08 | .657 | 0.01 |
| Computing XP | -0.35 | .056 | 0.12 |
| Smartphone XP | -0.40 | .027* | 0.16 |
| Gaming XP | -0.44 | .024* | 0.17 |
| MSA-Child | 0.40 | .031* | 0.16 |
| MSB-Adult | 0.66 | <.001*** | 0.39 |
| MSSQ Total | 0.47 | .010** | 0.22 |
| Pupil – CSQ-VR | -0.35 | .056 | 0.15 |
| Pupil – Ride | -0.38 | .036* | 0.16 |

XP = Experience; MS = Motion Sickness; MSSQ = Motion Sickness Susceptibility Questionnaire; Pupil – CSQ-VR = Pupil Size While Reading CSQ-VR Questions; Pupil Ride = Pupil Size During Ride; * p ≤ .05, ** p ≤ .01, *** p ≤ .001; & unstandardized beta

Table 3 provides insights specific to nausea symptoms. Once again, in line with **H1**, pupil size, both during CSQ-VR reading and the VR ride, was a significant predictor of nausea symptomatology. Notably, in agreement with **H2**, MSB-Adult stands out as the most potent predictor. The child motion sickness score (MSA-Child) and the total motion sickness susceptibility (MSSQ Total) were also significant, hinting that one's propensity to motion sickness earlier in life could have repercussions in a virtual environment. Finally, in support of **H3**, experience in using smartphones and playing videogames were significant predictors of nausea symptoms' intensity.

**Table 3.** Single Predictor Models for Nausea Symptoms

| Predictor | β coefficient | p-value | $R^2$ |
|---|---|---|---|
| Sex (Male) | -0.65& | .076 | 0.11 |
| Education | 0.13 | .484 | 0.02 |
| Age | 0.04 | .851 | 0.00 |
| Computing XP | -0.34 | .064 | 0.12 |
| Smartphone XP | -0.37 | .045* | 014 |
| Gaming XP | -0.38 | .041* | 0.14 |
| MSA-Child | 0.44 | .014* | 0.20 |
| MSB-Adult | 0.63 | <.001*** | 0.39 |
| MSSQ Total | 0.54 | .002** | 0.29 |
| Pupil – CSQ-VR | -0.38 | .041* | 0.14 |
| Pupil – Ride | -0.36 | .043* | 0.14 |

XP = Experience; MS = Motion Sickness; MSSQ = Motion Sickness Susceptibility Questionnaire; Pupil – CSQ-VR = Pupil Size While Reading CSQ-VR Questions; Pupil Ride = Pupil Size During Ride; * p ≤ .05, ** p ≤ .01, *** p ≤ .001; & unstandardized beta

In Table 4, the focus shifts to vestibular symptoms. While some predictors overlap with those for nausea, in disagreement with **H1**, the pupil size did not significantly predict vestibular symptoms. However, the MSB-Adult was found to be a robust predictor of nausea, further confirming **H2**. Finally, it is intriguing to see that smartphone and gaming experience are significant predictors of vestibular symptomatology, which agrees with **H3** and implies a potential link between technological experience and experiencing vestibular symptoms in a VR.

**Table 4.** Single Predictor Models for Vestibular Symptoms

| Predictor | β coefficient | p-value | R² |
|---|---|---|---|
| Sex (Male) | -0.61& | .100 | 0.09 |
| Education | 0.24 | .203 | 0.06 |
| Age | 0.00 | .980 | 0.00 |
| Computing XP | -0.22 | .238 | 0.05 |
| Smartphone XP | -0.31 | .047* | 0.13 |
| Gaming XP | -0.40 | .029* | 0.16 |
| MSA-Child | 0.20 | .290 | 0.04 |
| MSB-Adult | 0.53 | .003** | 0.28 |
| MSSQ Total | 0.28 | .137 | 0.08 |
| Pupil – CSQ-VR | -0.25 | .182 | 0.06 |
| Pupil – Ride | -0.24 | .208 | 0.06 |

XP = Experience; MS = Motion Sickness; MSSQ = Motion Sickness Susceptibility Questionnaire; Pupil – CSQ-VR = Pupil Size While Reading CSQ-VR Questions; Pupil Ride = Pupil Size During Ride; * p ≤ .05, ** p ≤ .01, *** p ≤ .001; & unstandardized beta

Table 5 centers on oculomotor symptoms. Unlike previous findings and discrepantly to **H1-H3**, no single predictor emerges as significant. The table largely communicates that more conventional metrics (age, education, tech experience) don't have significant associations with oculomotor symptoms in VR. It highlights the complexities of predicting these specific symptoms.

**Table 5.** Single Predictor Models for Oculomotor Symptoms

| Predictor | β coefficient | p-value | R² |
|---|---|---|---|
| Sex (Male) | -0.26& | .486 | 0.02 |
| Education | -0.14 | .468 | 0.02 |
| Age | -0.23 | .213 | 0.06 |
| Computing XP | -0.25 | .179 | 0.06 |
| Smartphone XP | -0.22 | .249 | 0.05 |
| Gaming XP | -0.18 | .336 | 0.03 |
| MSA-Child | 0.21 | .265 | 0.04 |
| MSB-Adult | 0.27 | .152 | 0.07 |
| MSSQ Total | 0.23 | .221 | 0.05 |
| Pupil – CSQ-VR | -0.11 | .569 | 0.01 |
| Pupil – Ride | -0.28 | .138 | 0.08 |

XP = Experience; MS = Motion Sickness; MSSQ = Motion Sickness Susceptibility Questionnaire; Pupil – CSQ-VR = Pupil Size While Reading CSQ-VR Questions; Pupil Ride = Pupil Size During Ride; * p ≤ .05, ** p ≤ .01, *** p ≤ .001; & unstandardized beta

Table 6 aggregates the best models for predicting various aspects of cybersickness. This table reemphasizes the overarching role of MSB-Adult in determining cybersickness, nausea, and vestibular symptoms. This connotes that the susceptibility to experiencing motion sickness as an adult is robustly associated with experiencing visually induced cybersickness symptoms in VR. Interestingly, no predictors were identified for oculomotor symptoms, indicating a potential gap in our understanding or the need for more refined measures.

**Table 6.** Best Models for Predicting Cybersickness

| Predicted | Predictors | β coefficient | p-value(β) | R² |
|---|---|---|---|---|
| CSQ-VR – Total | MSB-Adult | 0.66 | <.001*** | 0.39 |
| CSQ-VR – Nausea | MSB-Adult | 0.63 | <.001*** | 0.39 |
| CSQ-VR – Vestibular | MSB-Adult | 0.53 | .003** | 0.28 |
| CSQ-VR – Oculomotor | Null Model | – | – | – |

CSQ-VR = Cybersickness in Virtual reality Questionnaire; MS = Motion Sickness;
* p ≤ .05, ** p ≤ .01, *** p ≤ .001

In essence, these results underscore the multifaceted nature of cybersickness and its determinants. While some predictors like MSB-Adult consistently emerge as influential, others show symptom-specific associations. Furthermore, they accentuate the importance of not only considering the user's history and demographics but also real-time metrics like pupil size in understanding their VR experience.

### 3.2. Mixed Model Regression Analyses: Effects on Cognitive & Motor Performance

Mixed linear regression model analyses were carried out to determine and evaluate the significant predictors of performance on cognitive and psychomotor skills tasks. Tables 7 and 8 delve into the predictors for verbal short-term and working memory, respectively. In disagreement with **H4**, the critical observation here is that none of the predictors were significant predictors of verbal short-term and working memory. This might suggest that verbal memory is less susceptible to variations in these predictors or that other unmeasured factors may have a more substantial impact.

**Table 7.** Single Predictor Models for Verbal Short-Term Memory

| Predictor | β coefficient | p-value | R²(Fixed Effects / Overall) |
|---|---|---|---|
| Sex (Male) | 0.35⁎ | .249 | 0.03 / 0.52 |
| Education | -0.13 | .411 | 0.02 / 0.52 |
| Age | 0.09 | .534 | 0.01 / 0.52 |
| Computing XP | 0.10 | .496 | 0.01 / 0.52 |
| Smartphone XP | 0.29 | .153 | 0.05 / 0.52 |
| Gaming XP | 0.10 | .525 | 0.01 / 0.52 |
| CSQ-VR – Total | -0.12 | .244 | 0.02 / 0.51 |
| CSQ-VR – Nausea | -0.01 | .965 | 0.01 / 0.50 |
| CSQ-VR – Vestibular | -0.17 | .105 | 0.03 / 0.52 |
| CSQ-VR – Oculomotor | -0.15 | .173 | 0.03 / 0.54 |

CSQ-VR = Cybersickness in Virtual reality Questionnaire; XP = Experience;
* p ≤ .05, ** p ≤ .01, *** p ≤ .001; ⁎ unstandardized beta

**Table 8.** Single Predictor Models for Verbal Working Memory

| Predictor | β coefficient | p-value | R²(Fixed Effects / Overall) |
|---|---|---|---|
| Sex (Male) | 0.27⁎ | .404 | 0.02 / 0.54 |
| Education | 0.09 | . 592 | 0.01 / 0.54 |
| Age | 0.15 | .351 | 0.02 / 0.54 |
| Computing XP | -0.02 | .911 | 0.01 / 0.54 |
| Smartphone XP | 0.10 | .668 | 0.01 / 0.54 |
| Gaming XP | 0.07 | .679 | 0.01 / 0.54 |
| CSQ-VR – Total | -0.03 | .798 | 0.01 / 0.53 |
| CSQ-VR – Nausea | -0.02 | .862 | 0.01 / 0.52 |
| CSQ-VR – Vestibular | -0.05 | .669 | 0.01 / 0.53 |
| CSQ-VR – Oculomotor | -0.01 | .941 | 0.01 / 0.53 |

CSQ-VR = Cybersickness in Virtual reality Questionnaire; XP = Experience;

* p ≤ .05, ** p ≤ .01, *** p ≤ .001; & unstandardized beta

Table 9 elucidates predictors for visuospatial short-term memory. In contrast with **H4**, none of the cybersickness measurements was found to be a significant predictor of performance. However, it shows some compelling findings. Sex, computing experience, and gaming experience were shown as significant predictors of visuospatial short-term memory. Gaming experience, in particular, accounts for a substantial 19% of the variance.

**Table 9.** Single Predictor Models for Visuospatial Short-Term Memory

| Predictor | β coefficient | p-value | R²(Fixed Effects / Overall) |
|---|---|---|---|
| Sex (Male) | 0.83& | .006** | 0.17 / 0.56 |
| Education | -0.13 | .468 | 0.07 / 0.57 |
| Age | 0.15 | .363 | 0.02 / 0.57 |
| Computing XP | 0.34 | .029* | 0.11 / 0.57 |
| Smartphone XP | 0.12 | .476 | 0.01 / 0.57 |
| Gaming XP | 0.46 | .003** | 0.19 / 0.56 |
| CSQ-VR – Total | -0.07 | .538 | 0.01 / 0.54 |
| CSQ-VR – Nausea | -0.06 | .541 | 0.01 / 0.54 |
| CSQ-VR – Vestibular | -0.04 | .694 | 0.01 / 0.54 |
| CSQ-VR – Oculomotor | -0.04 | .760 | 0.01 / 0.55 |

CSQ-VR = Cybersickness in Virtual reality Questionnaire; XP = Experience;
* p ≤ .05, ** p ≤ .01, *** p ≤ .001; & unstandardized beta

Table 10 examines predictors for visuospatial working memory. Supporting **H4**, the vestibular symptoms intensity was found to be a significant predictor of visuospatial working memory, albeit other cybersickness metrics did not substantially predict it. Furthermore, sex and gaming experience stand out as significant predictors, indicating that gaming might shape how individuals process and manipulate visual-spatial information.

**Table 10.** Single Predictor Models for Visuospatial Working Memory

| Predictor | β coefficient | p-value | R²(Fixed Effects / Overall) |
|---|---|---|---|
| Sex (Male) | 0.82& | .013* | 0.14 / 0.65 |
| Education | -0.12 | . 519 | 0.01 / 0.65 |
| Age | 0.20 | .233 | 0.04 / 0.65 |
| Computing XP | 0.21 | .221 | 0.04 / 0.65 |
| Smartphone XP | 0.30 | .186 | 0.05 / 0.65 |
| Gaming XP | 0.40 | .015* | 0.14 / 0.65 |
| CSQ-VR – Total | -0.19 | .056 | 0.04 / 0.63 |
| CSQ-VR – Nausea | -0.12 | .209 | 0.02 / 0.63 |
| CSQ-VR – Vestibular | -0.23 | .022* | 0.05 / 0.66 |
| CSQ-VR – Oculomotor | -0.16 | .141 | 0.03 / 0.63 |

CSQ-VR = Cybersickness in Virtual reality Questionnaire; XP = Experience;
* p ≤ .05, ** p ≤ .01, *** p ≤ .001; & unstandardized beta

Table 11 on attentional time displays that CSQ-VR—specifically the oculomotor component—has a notable relationship with attentional time. The positive β coefficient suggests a direct correlation, meaning that as cybersickness symptoms increase, attentional time might also increase.

**Table 11.** Single Predictor Models for Attentional Time

| Predictor | β coefficient | p-value | R²(Fixed Effects / Overall) |
|---|---|---|---|
| Sex (Male) | -0.06& | .837 | 0.01 / 0.33 |
| Education | -0.14 | .367 | 0.02 / 0.33 |
| Age | -0.16 | .276 | 0.03 / 0.33 |
| Computing XP | -0.28 | .060 | 0.07 / 0.33 |
| Smartphone XP | -0.17 | .249 | 0.03 / 0.33 |
| Gaming XP | -0.25 | .093 | 0.06 / 0.33 |
| CSQ-VR – Total | 0.25 | .044* | 0.06 / 0.34 |
| CSQ-VR – Nausea | 0.21 | .078 | 0.05 / 0.32 |
| CSQ-VR – Vestibular | 0.16 | .192 | 0.03 / 0.32 |
| CSQ-VR – Oculomotor | 0.28 | .026* | 0.08 / 0.36 |

CSQ-VR = Cybersickness in Virtual reality Questionnaire; XP = Experience;

\* p ≤ .05, \*\* p ≤ .01, \*\*\* p ≤ .001; & unstandardized beta

Table 12 targets motor time. Here, gaming experience and CSQ-VR – Nausea are prominent predictors. This implies that one's gaming experience might influence motor response times, and individuals experiencing nausea-related cybersickness symptoms might exhibit changes in their motor time.

**Table 12.** Single Predictor Models for Motor Time

| Predictor | β coefficient | p-value | R²(Fixed Effects / Overall) |
|---|---|---|---|
| Sex (Male) | -0.59& | .075 | 0.08 / 0.69 |
| Education | 0.18 | .298 | 0.03 / 0.69 |
| Age | -0.03 | .874 | 0.01 / 0.69 |
| Computing XP | -0.14 | .417 | 0.02 / 0.69 |
| Smartphone XP | -0.14 | .410 | 0.02 / 0.69 |
| Gaming XP | -0.38 | .018* | 0.14 / 0.69 |
| CSQ-VR – Total | 0.18 | .067 | 0.03 / 0.67 |
| CSQ-VR – Nausea | 0.20 | .030* | 0.04 / 0.67 |
| CSQ-VR – Vestibular | 0.14 | .145 | 0.02 / 0.67 |
| CSQ-VR – Oculomotor | 0.09 | .368 | 0.01 / 0.67 |

CSQ-VR = Cybersickness in Virtual reality Questionnaire; XP = Experience;

\* p ≤ .05, \*\* p ≤ .01, \*\*\* p ≤ .001; & unstandardized beta

In Table 13, the focus is on reaction time (i.e., psychomotor skills). In full support of **H4**, every cybersickness metric was a significant predictor of psychomotor skills, postulating that overall cybersickness, nausea, vestibular, and oculomotor symptoms play a role in determining reaction times. Interestingly, gaming experience also emerges as a significant negative predictor, likely hinting that frequent gamers might have quicker reaction times.

**Table 13.** Single Predictor Models for Reaction Time

| Predictor | β coefficient | p-value | R²(Fixed Effects / Overall) |
|---|---|---|---|
| Sex (Male) | -0.65& | .054 | 0.10 / 0.75 |
| Education | 0.10 | . 590 | 0.01 / 0.75 |
| Age | -0.07 | .691 | 0.01 / 0.75 |
| Computing XP | -0.24 | .157 | 0.06 / 0.76 |
| Smartphone XP | -0.20 | .239 | 0.04 / 0.75 |
| Gaming XP | -0.41 | .013* | 0.16 / 0.75 |
| CSQ-VR – Total | 0.25 | .004** | 0.06 / 0.76 |
| CSQ-VR – Nausea | 0.22 | .009** | 0.04 / 0.74 |
| CSQ-VR – Vestibular | 0.19 | .028* | 0.04 / 0.75 |
| CSQ-VR – Oculomotor | 0.22 | .020* | 0.05 / 0.76 |

CSQ-VR = Cybersickness in Virtual reality Questionnaire; XP = Experience;

\* p ≤ .05, \*\* p ≤ .01, \*\*\* p ≤ .001; & unstandardized beta

Lastly, Table 14 consolidates the most impactful predictors for various cognitive and motor skills. A recurring theme here is the influence of gaming experience on cognitive and motor skills, emphasizing its potential cognitive benefits or the development of specific skills associated with gaming. Interestingly, overall cybersickness, vestibular and oculomotor components, were also included in the best models, postulating the negative effects of cybersickness on cognitive functioning and psychomotor skills.

**Table 14.** Best Models for Predicting Cognitive & Motor Skills

| Predicted | Predictors | β coefficient | p-value(β) | R²(Fixed Effects / Overall) |
|---|---|---|---|---|
| DST - Forward | Null Model | – | – | – |
| DST - Backward | Null Model | – | – | – |
| CBT - Forward | Gaming XP | 0.46 | .003** | 0.19 / 0.56 |
| CBT - Backward | Gaming XP | 0.36 | .034* | 0.20 / 0.68 |
| | CSQ-VR – Vestibular | -0.21 | .025* | |
| DLRT – AT | CSQ-VR – Oculomotor | 0.28 | .026* | 0.08 / 0.36 |
| DLRT – MT | Gaming XP | -0.36 | .023* | 0.18 / 0.69 |
| | CSQ-VR – Nausea | 0.19 | .049* | |
| DLRT – RT | Gaming XP | -0.35 | .028* | 0.21 / 0.77 |
| | CSQ-VR – Total | 0.23 | .008** | |

CSQ-VR = Cybersickness in Virtual reality Questionnaire; XP = Experience; DST = Digit Span Test; CBT = Corsi Block Test; DLRT = Deary-Liewald Reaction Time Test; RT = Reaction Time; AT = Attention Time; MT = Motor Time; * p ≤ .05, ** p ≤ .01, *** p ≤ .001

In summary, these results underscore the potential cognitive and motor influences of digital experiences like gaming. They also hint at the intertwined nature of cybersickness symptoms and cognitive/motor skills, suggesting that our experience in virtual environments can have multifaceted impacts on cognitive functioning.

### 3.3. Comparison of Cybersickness Between During and After Exposure to Virtual Reality

Paired sample t-test analyses were performed to examine if there was a difference in terms of cybersickness symptomatology between during VR immersion and after exposure to VR (i.e., immediately after the removal of VR equipment). Figure 4 illustrates the z-scores representing the intensity of overall and various cybersickness symptoms experienced by participants during their immersion in VR and after their exposure to VR (i.e., after the removal of the VR headset). In line with **H5**, the overall cybersickness intensity was found to have a significant and large decrease after the removal of the VR headset (i.e., the re-adaptation to the physical world), *t(29)=3.59, p<.001, Hedges g = 0.64*. Furthermore, other cybersickness symptoms showed significant decreases after removing the VR headset and transitioning from the virtual to the physical environment, further supporting the **H5**.

For vestibular symptoms, after exposure to VR (Post-VR), a noticeable shift towards negative z-scores was identified, indicating that the experience of vestibular symptoms was less intense. This difference between the two stages was confirmed to be statistically significant and large, *t(29)=2.74, p = .010, Hedges g = 0.49*. Regarding nausea, a slightly broader spread of z-scores, predominantly on the negative side, was detected. This decrease in nausea symptoms' intensity was deemed significant and large, *t(29)=3.12, p<.001, Hedges g = 0.56*. Finally, for oculomotor symptoms, after exposure to VR, the distribution pattern of the z-scores remained largely unchanged, *t(29)=1.22, p = .230, Hedges g = 0.22*, indicating that the removal of the headset and re-adaptation to the physical world had a non-significant effect on oculomotor symptoms' intensity. To summarize, of the three cybersickness symptom categories examined, the vestibular and nausea symptoms exhibited a statistically significant decrease after the headset removal (i.e., post-VR exposure). It was determined, however, that oculomotor symptoms' intensity remained relatively stable after the exposure to VR.

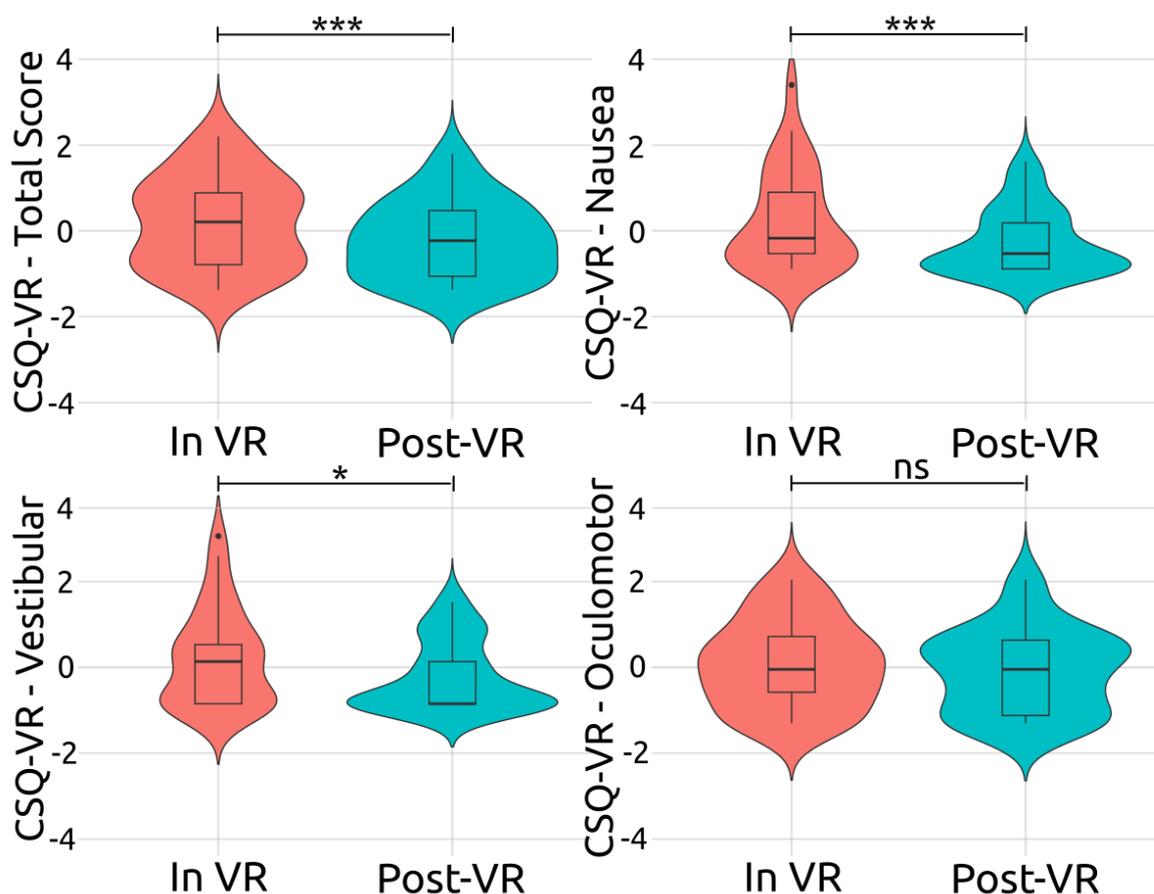

**Figure 4.** Comparison of Cybersickness' Intensity Between During and After Exposure to Virtual Reality. Note: CSQ-VR = Cybersickness in Virtual reality Questionnaire; In VR = During Immersion; Post-VR = After Exposure to VR; ns = non-significant, * p ≤ .05, ** p ≤ .01, *** p ≤ .001

## 4. Discussion

Considering that VR is implemented in educational, professional, research, and clinical settings, the present study aimed to examine cybersickness symptoms during immersion in VR. The study explored the role of several factors pertinent to individual differences, such as motion sickness susceptibility, experience in playing videogames, experience in using computers, sex, and age, in experiencing cybersickness symptomatology and intensity. For predicting cybersickness, overall and symptoms intensity, eye-tracking metrics, such as pupil dilation, were also considered. Furthermore, the study aimed to determine the effects of cybersickness symptoms on cognitive functions and motor skills. Finally, given that users experience a readjustment to physical space, while removing the VR HMD and transitioning from the virtual to the physical environment, the current study also examined the differences in cybersickness intensity during and after immersion. A comprehensive discussion is offered by integrating findings from the current study with the insights from the reviewed literature.

### 4.1. Pupil Dilation as a Biomarker of Cybersickness

Pupil dilation has been previously seen as a biomarker of affective state, where a bigger size indicates a positive (e.g., joy) and a smaller size indicates a negative affective state (e.g., fear) [102]. Our results demonstrated a comparable pattern in cybersickness, where more intense cybersickness induced a smaller diameter (i.e., more negative affective state), and less intense a bigger diameter of the pupils (i.e., more positive affective state). In VR, pupil size has been previously incorporated into a deep fusion model to predict cybersickness [103]. However, this previous study did not assess its association, predictive capacity, and role within this model, making it challenging to determine if pupil size acts as a biomarker for cybersickness. However, in our previous studies, evidence was offered that pupil dilation while reading the questions of CSQ-VR substantially predicts cybersickness intensity [49], [53]. In line with all the studies mentioned above, pupil dilation was found to be a significant predictor of cybersickness in this study. Nevertheless, the pupil dilation was measured during reading CSQ-VR questions, as well as during the ride. The pupil size during the ride was deemed a significant predictor of overall cybersickness and nausea symptoms, while the pupil size during responding to CSQ-VR significantly predicted only the overall cybersickness. Also, for overall cybersickness intensity, pupil dilation during the ride was a substantially better predictor than pupil dilation during CSQ-VR. This suggests that pupil size is a more reliable biomarker during the triggering and experience of cybersickness than post-exposure to stimuli eliciting symptomatology.

### 4.2. Modulators of Cybersickness: Sex, Smartphone Experience, and Videogame Experience

None of the demographics appeared to be a significant predictor of cybersickness intensity. Notably, age was not a significant predictor of cybersickness. Given the younger demographic of participants, it is worth investigating these effects in a broader age range. Also, experience in using computers failed to predict cybersickness. Finally, in disagreement with previous studies [73], [104], sex (i.e., male, female) did not predict cybersickness intensity and symptoms, postulating that cybersickness is not more frequent or intense in either sex. Interestingly, Stanney et al. [73] proposed that the variations stemmed from the VR headset's InterPupillary Distance (IPD). In their subsequent experiment, no disparities between the two sexes were observed when participants deemed the IPD agreeable. In our research, the HTC Vive Pro Eye was employed, known for its universal comfort. Each participant underwent an eye-tracking calibration to fine-tune the IPD, ensuring its appropriateness and comfort. Also, in our previous study [49], the differences between sexes in terms of cybersickness were eliminated when we con-trolled for gaming experience (i.e., male and female users with the same experience in playing videogames). In this study, a balanced sample was attempted for having both female and male participants with a comparable (e.g., high/low) gaming experience. Therefore, the calibration of the IPD via eye-tracking, as well as the balance in terms of gaming experience among participants, may explain the non-significant effect of sex on experiencing cybersickness. In combination, the findings of this study and our previous study [49], along with the findings of Stanney et al., [73] suggest that the sex/gender of the participant does not modulate the intensity of the cybersickness symptomatology.

On the contrary, experience in playing videogames and experience in using apps on the smartphone for improving everyday functionality significantly predicted the intensity of perceived cybersickness symptoms. Notably, our study is the first to show that experience with smartphones may be a predictor of cybersickness. Specifically, higher experience with smartphones was associated with a substantially lower intensity of cybersickness symptomatology. Interestingly, the usage of smartphones for performing various tasks (e.g., surfing on the internet, sending emails, writing documents, editing photos and videos, etc.), which traditionally were performed on a computer, nowadays is significantly higher compared to computers/laptops' usage for performing the same tasks [105]. Note that visually induced cybersickness can be elicited by exposure to any screen, including smartphones [106]–[108]. Also, exposure to and experience with using tech mediums with screens is associated with cultivating tolerance towards experiencing cybersickness symptomatology by adapting to spatial requirements and motion [32]. Thus, the finding of this study regarding smartphone experience for developing resilience to cybersickness is aligned with the relevant literature.

In the same direction, the current study showed that higher experience with playing videogames postulates a higher resilience to cybersickness. This finding corroborates with the findings of our previous studies [49], [53], a large scale study by Weech et al., [77], as well as other studies on cybersickness [109] and simulator sickness [110]. Therefore, a more significant gaming background seems to act as a protective factor against cybersickness, whereas a limited one might heighten vulnerability. Notably, similar to our previous studies on cybersickness [49], [53], in this study, experience in playing videogames considered both proficiency and frequency, which appears as a refined measure of gaming experience. However, the videogames are clustered under several diverse genres (e.g., action, first-person shooting, role-playing, etc.), and each one may modulate the physiological and biochemical state [111], as well as cognitive functioning and brain structure [112], in a significantly different way. Thus, further research is required for examining the effects of gaming on cybersickness, by considering experience in playing games of each genre as distinct metrics.

### 4.3. Cybersickness & Susceptibility to Motion Sickness

Susceptibility to motion sickness during adulthood (i.e., MSB-Adult score of MSSQ) emerged as the best predictor of cybersickness intensity. Except for oculomotor symptoms, the MSB-Adult was a significant predictor of overall cybersickness, nausea symptoms, and vestibular symptoms. Notably, the single predictor models with MSB-Adult (i.e., having only MSB-Adult as a predictor) were found to be the best regression models for predicting the respective scores of CSQ-VR. Considering that motion sickness and cybersickness induced by vection share common characteristics, such as motion cues acting as elicitors of sickness, it comes with no surprise that findings of previous studies postulated that visually induced cybersickness and motion sickness demonstrate similar patterns of susceptibility among individuals [78], [79]. The findings of this study agree with the findings of the aforementioned studies, since the MSB-Adult was the most prominent predictor of overall cybersickness, as well as nausea and vestibular symptoms.

However, in our previous study MSSQ scores failed to predict cybersickness intensity and symptoms [49]. Nonetheless, in this previous study, the MSSQ was also used to exclude participants who demonstrated high susceptibility to motion sickness. This may explain why the MSSQ scores were not identified as significant predictors in our previous study. Finally, taken together, the MSSQ scores, especially MSB-Adult, may be used to identify individuals who are prone to experience a vection induced cybersickness. However, the Visual Induced MSSQ (VIMSSQ) has been recently developed and validated, which is specific to exposure to screens (e.g., computers, tablets, and VR HMDs) [108], [113]. Also, it should be noted that vection is only one of the reasons for experiencing cybersickness in VR [31], [35], [36], [53]. Thus, further research is required to examine whether MSSQ and/or VIMSSQ scores may predict cybersickness induced by other factors such as low latency or latency fluctuations, non-ergonomic navigation, and low-quality graphics.

### 4.4. Cybersickness Effects on Verbal Short-Term Memory and Working Memory

The study of Dahlman et al. [84] posited a direct negative effect of motion sickness on verbal working memory. Also, significant negative effects of cybersickness on verbal working memory were observed in our previous study [49]. Discrepantly to these previous studies, cybersickness was seen to affect significantly neither verbal short-term memory nor verbal working memory. Regarding short-term and working memory, there is consensus that they are two different cognitive constructs stemming from the activation of diverse brain regions, where the former requires substantially fewer cognitive resources than the latter [114], [115]. Hence, the short-term memory may not appear to decrease due to low difficulty. Furthermore, the study of Dahlman et al. [84] was on motion sickness. While motion sickness and cybersickness share some similarities, they are substantially different in terms of symptoms frequency and intensity [29], [30]. Thus, this difference between cybersickness and motion sickness may explain the disagreement between the findings of this study and the findings of Dahlman and collaborators' study.

Furthermore, in our previous study, only working memory was examined [49]. Also, the order of the tasks was not counterbalanced, where the verbal working memory task was always first to be performed after exposure to linear and angular accelerations. Finally, the size effect of the performance decrease on the verbal working memory task was small [49]. These limitations of our previous study may thus explain the discrepancy between the findings of the two studies. Nonetheless, given that cybersickness effects are transient and of relatively short duration [52], if performing a task, immediately after exposure to stimuli inducing cybersickness, does indeed have an impact, then this poses a severe methodological consideration that should be further explored in future studies. However, in this study, in line with the design of the original tests (see [91], [93]), short-term memory tasks always precede working memory tasks. Thus, if the order had a significant impact, cybersickness effects should have also been observed on short-term memory tasks.

### 4.5. Cybersickness Effects on Visuospatial Short-Term Memory and Working Memory

Cybersickness was found to have a significant negative effect on visuospatial working memory. This finding aligns with the findings of the study of Mittelstaedt et al. [59], where performance on the visuospatial working memory task was substantially decreased. The two studies hence postulate that cybersickness has indeed a negative impact on visuospatial working memory. However, again, there was no effect on visuospatial short-term memory. Considering that administration of the CBT tasks, forward (short-term memory) and backward (working memory) recall tasks, should be in this order (i.e., forward recall and then backward recall) [93], [94], the absence of an effect on short-term memory and a significant effect on working memory, further supports that order of the tasks does not have an impact on observing effects of cybersickness on cognitive performance. Furthermore, similar to verbal short-term and working memory, it is widely agreed that visuospatial short-term memory and working memory are dis-tinct cognitive structures facilitated by different brain structures, with the former re-quiring fewer cognitive resources than the latter [115], [116]. This difference between the two explains why visuospatial short-term memory was left intact by cybersickness while visuospatial working memory was substantially decreased.

Moreover, in this study, vestibular symptomatology, which postulates a transient dysfunction of the vestibular system, was found to have a significant negative impact on visuospatial working memory. Visuospatial working memory functioning pertains to the processing of visuospatial information [115], [116]. The vestibular system has been suggested to have an important implication in visuospatial cognitive functioning [117]. Thus, the decrease of visuospatial working memory by predominantly vestibular symptomatology that was observed in this study further supports the importance of the vestibular system to visuospatial information processing. Nevertheless, the best model for predicting visuospatial working memory also included gaming experience, which revealed a positive effect on working memory. Notably, gaming experience was also included in the best model of visuospatial short-term memory. These findings are in line with the relevant literature, which suggests that gamers have enhanced short-term and working memory abilities [118]. In the investigation of cybersickness, these findings postulate that gaming experience should always be considered (e.g., as a covariate or an additional factor) when examining the effects of cybersickness on cognition.

*4.6. Cybersickness Effects on Psychomotor Skills: Reaction, Attention, and Motor Speed*

Psychomotor skills, such as attentional speed, motor speed, and overall reaction time were found to be substantially negatively affected by cybersickness symptomatology and intensity. In this study, the VR version of DLRT was implemented, which in contrast with the traditional version that produces a single score (i.e., reaction time), the DLRT-VR produces three metrics corresponding to attentional speed, motor speed, and overall reaction time. Since the previous studies on cybersickness (e.g., [34], [59], [60]) used the traditional version, and assessed participants after immersion, the current study may further decipher the effects of cybersickness on psychomotor skills. The observed significant deceleration of overall reaction speed is in line with our [49], [53] and other previous studies [34], [59], [60]. In conjunction, the current and previous findings thus offer robust evidence that cybersickness substantially compromised psychomotor skills. However, the experience in playing videogames was also included in the best model, where a significant positive effect on reaction time (i.e., acceleration of reaction speed) was detected. This finding aligns with the previous literature pertaining to the effects of gaming experience on psychomotor speed [119]–[122]. In the context of cybersickness, this outcome postulates that the gaming experience has to be considered (e.g., as a covariate) for effectively examining the effects of cybersickness on psychomotor skills.

While the effects on overall reaction speed are well established by the current and previous findings, the cybersickness effects on the components of psychomotor skills (i.e., attentional and motor speed) still need to be investigated in depth. In this study, the attentional speed was found to be significantly decelerated by the oculomotor symptoms' intensity. This outcome is in agreement with previous studies that revealed a significant negative effect on attentional processing speed [59], [85]. However, these previous studies attributed the deceleration of attentional speed to overall cybersickness. In the current study, while both overall cybersickness intensity and oculomotor symptoms' intensity were deemed significant predictors of attentional speed, only the oculomotor symptomatology was incorporated in the respective best model. This outcome is in line with the established understanding that the oculomotor system is essential for facilitating visual attention functioning, especially for orienting attention [123], [124]. Hence, our findings connote that a transient dysfunction of the oculomotor system (e.g., eye fatigue or strain) substantially compromises attentional speed.

However, the deceleration of the motor speed was found to be predominantly attributed to nausea symptomatology. This finding is aligned with the current understanding of the negative effects of nausea on motor coordination and skills due to a modulation of the activation of sensorimotor brain regions [125], [126]. Nevertheless, the gaming experience was also included in the best predictive model of motor speed, where a significant acceleration of motor speed was observed due to higher gaming experience. This aligns with the relevant literature that postulates that the gaming experience promotes an enhanced motor speed, especially for fine motor functions [127], [128]. Thus, this finding suggests that gaming experience should be considered in the examination of cybersickness effects on motor speed. In summary, gaming experience appears central to enhancing psychomotor speed, while cybersickness substantially decelerates overall reaction speed. Regarding the components of psychomotor skills, oculomotor and nausea symptoms' intensity significantly decelerates attentional and motor speed respectively.

*4.7. Cybersickness Symptoms and Intensity During and After Immersion*

The intensities of overall cybersickness, nausea symptoms, and vestibular symptoms were substantially decreased after immersion (i.e., immediately after removing the headset). However, oculomotor symptoms did not decrease, which suggests that they may have a more lasting impact. This appears similar to simulator sickness, where oculomotor symptoms like eye-strain persist long after the exposure to the simulator [129]. In contrast, vestibular and nausea symptoms are the most frequent and predominant in cybersickness [29], [61], [130]. Thus, these findings draw a stark contrast between the experience of cybersickness during immersion in a virtual environment and after immersion. This agrees with the suggestion that the human body and mind strive to readjust to the physical environment immediately after the removal of the VR headset [86].

Based on our results, after the exposure to stimuli inducing cybersickness, and during this transitory period of readapting to the physical space and body, the cybersickness intensity and symptoms substantially fade away. This also agrees with the current predominant understanding that cybersickness symptoms and effects are transient [52]. However, there is no consensus concerning how long cybersickness symptoms may persist, with some previous reviews suggesting that they may last for up to 12 hours after exposure to VR [31], [131]. However, our results postulate that the period of experiencing a substantial alleviation of cybersickness intensity commences during immersion (i.e., after exposure to stimuli inducing cybersickness, such as linear and angular accelerations) and have attained a significant decrease after immersion, during the readaptation to the physical body and environment. Hence, this natural process counteracts or alleviates some of the dissonances experienced in the virtual environment during immersion.

Furthermore, these findings posit fundamental ramifications for the methodology used in VR research on cybersickness. To the best of our knowledge, except for the current and our previous studies [49], [53], the studies on cybersickness (e.g., [28], [45]–[48], [56]) or cybersickness effects on human physiology or cognitive functioning or psychomotor skills (e.g., [34], [59], [60], [64]–[66], [85], [130]) measure cybersickness intensity, symptomatology, and/or its effects, after immersion. Our findings indicate that this approach may only capture a part of the overall picture, or even worse, a substantially distorted picture.

If cybersickness symptoms subside or change in intensity immediately after VR exposure, then solely post-immersion evaluations could lead to unreliable conclusions. For instance, evaluations done post-immersion might underestimate the true intensity of symptoms experienced during VR exposure, as well as the effects of cybersickness on cognition and motor skills. From a research design standpoint, these insights underscore the importance of adopting a more temporally accurate approach. Hence, Hence, in studies attempting to attain a holistic view of cybersickness, researchers should endeavor to capture data at multiple points – before, during, and after VR exposure. On the other hand, studies attempting to examine cybersickness aftereffects or persistence of symptomatology may perform the examination after immersion. Lastly, it is crucial for studies striving to assess cybersickness intensity and symptomatology, as well as its effects on physiology (e.g., autonomic responses such as heart rate and temperature), cognition, and motor skills to perform their examination during the VR immersion.

### 4.8. Limitations and Future Studies

This study also has some limitations that should be considered. While allowing the conduction of the required statistical analyses, the sample size was relatively small. Also, the sample consisted predominantly of young adults aged 20-45 years old. A future study should incorporate a larger and/or more age-diverse sample (e.g., considering adolescents and/or older adults). Moreover, this study utilized the MSSQ and did not consider the VIMSSQ, which is specific to vection elicited by screen-based mediums. Using both in a future study may assist with deciphering whether generic motion sickness susceptibility or specific to vection is an indicator for experiencing cybersickness symptomatology and intensity. Also, the overall gaming experience was used for investigating its effects on cybersickness, cognitive functioning, and psychomotor skills. Since each genre of video-games offers diverse content that may stimulate distinct physiological and cognitive aspects, future studies should attempt to examine the effects of gaming experience by genre.

Moreover, this study explored the cybersickness intensity and symptoms in VR that were induced by vection. While vection is indeed one of the main reasons for cybersickness, several factors (e.g., low latency or latency fluctuations, non-ergonomic navigation, and low-quality graphics) may induce cyber-sickness. Future research should explore cybersickness intensity and symptomatology induced by each factor. The current study incorporated several cognitive and psychomotor tasks. Given that a task's characteristics may modulate cybersickness [132], future studies should either consider a single task or examine each task's effects on cybersickness. Finally, considering that significant differences were found during and post-immersion, future attempts should consider multiple assessment points of cybersickness to effectively scrutinize the cybersickness symptoms and intensity before, during (i.e., with several assessments), and after immersion.

## 5. Conclusions

This study comprehensively explored the modulators, symptomatology, and effects of cybersickness on an array of cognitive, physiological, and psychomotor functions. Notably, it was found that pupil dilation could be a valuable biomarker for cybersickness, reflecting the intensity of the experienced symptoms. Interestingly, demographic factors like sex and age did not significantly predict cybersickness. However, experience with videogames and smartphones emerged as protective factors, suggesting that familiarity with screen-based interactions could mitigate cybersickness symptoms. A particularly significant insight was the relationship between motion sickness susceptibility in adulthood and cybersickness intensity, suggesting an intertwined susceptibility pattern between these two phenomena. Furthermore, the research identified a selective impact of cybersickness when dissecting the effects on cognitive domains. While visuospatial working memory suffered from cybersickness, verbal short-term and working memory remained largely untouched. This distinction underscores the complex interplay between neural mechanisms and their susceptibility to cybersickness. One of the most crucial takeaways from this study was the marked difference in the experience and intensity of cybersickness during VR immersion compared to the post-immersion phase. Cybersickness symptomatology showed substantial changes post-immersion. This temporal difference poses significant methodological implications: if cybersickness evaluations are exclusively conducted post-immersion, they might not offer reliable results and conclusions. Therefore, for a more accurate understanding of cybersickness, future research endeavours should consider assessment points during immersion and/or post-immersion phases, respectively, to the research aims.

**Author Contributions:** Conceptualization, P.K.; methodology, P.K; software, P.K.; validation, A.P., P.R., and P.K; formal analysis, P.K.; investigation, A.P.; resources, P.R. and P.K; data curation, A.P. and P.K; writing—original draft preparation, A.P. and P.K; writing—review and editing, A.P., P.R., and P.K; visualization, P.K.; supervision, P.K.; project administration, P.R. and P.K; funding acquisition, A.P. and P.R. All authors have read and agreed to the published version of the manuscript.

**Funding:** This research received no external funding.

**Institutional Review Board Statement:** The study was conducted in accordance with the Declaration of Helsinki and approved by the Ethics Committee of Department of Psychology of the National and Kapodistrian University of Athens (796-14/07/2023).

**Informed Consent Statement:** Informed consent was obtained from all subjects involved in the study.

**Data Availability Statement:** The data presented in this study are available on request from the corresponding author. The data are not publicly available due to the ethical approval requirements.

**Conflicts of Interest:** The authors declare no conflict of interest.